\documentclass[%
 reprint,
superscriptaddress,
 amsmath,amssymb,
 aps,
prx,
]{revtex4-1}

\usepackage{graphicx}% Include figure files
\usepackage[usenames, dvipsnames]{color}
\usepackage{dcolumn}% Align table columns on decimal point
\usepackage{bm}% bold math
\usepackage{enumerate}  
\newcommand{\ket}[1]{| #1 \rangle}

\begin{document}

\preprint{APS/123-QED}

\title{Controlling chaos in the quantum regime using adaptive measurements}

\author{Jessica~K.~Eastman}
\email{jessica.eastman@anu.edu.au}
\affiliation{Department of Quantum Science, Research School of Physics and Engineering, The Australian National University, Canberra,  ACT 2601 Australia}
\affiliation{Centre for Quantum Computation and Communication Technology (Australian Research Council), Griffith University, Brisbane, Queensland 4111, Australia
} 
\affiliation{Centre for Quantum Dynamics, Griffith University, Brisbane, Queensland 4111, Australia}

 \author{Stuart~S.~Szigeti}%
\author{Joseph~J.~Hope}%
 \affiliation{Department of Quantum Science, Research School of Physics and Engineering, The Australian National University, Canberra,  ACT 2601 Australia}

\author{Andr\'e~R.~R.~Carvalho}
\affiliation{Centre for Quantum Dynamics, Griffith University, Brisbane, Queensland 4111, Australia}

\date{\today}

\begin{abstract}
The continuous monitoring of a quantum system strongly influences the emergence of chaotic dynamics near the transition from the quantum regime to the classical regime. 
Here we present a feedback control scheme that uses adaptive measurement techniques to control the degree of chaos in the driven-damped quantum Duffing oscillator. This control relies purely on the measurement backaction on the system, making it a uniquely quantum control, and is only possible due to the sensitivity of chaos to measurement. We quantify the effectiveness of our control by numerically computing the quantum Lyapunov exponent over a wide range of parameters. We demonstrate that adaptive measurement techniques can control the onset of chaos in the system, pushing the quantum-classical boundary further into the quantum regime.
\end{abstract}

\maketitle

Quantum systems possess uniquely nonclassical properties, such as coherence and entanglement, which can be manipulated for applications including quantum computation~\cite{Nielsen:2010,Harrow:2017}, quantum communication~\cite{Hammerer:2010,Liao:2017}, and quantum sensing~\cite{Giovannetti:2006, Pezze:2018}. Designing controls that do this is a diverse and productive area of ongoing research~\cite{Zhang:2017, Verstraete:2009, Santos:2012, Demkowicz-Dobrzanski:2014, Szigeti:2014b, Grimsmo:2015, Cramer:2016, Hirose:2016, Hosten:2016b, Szigeti:2017, Nolan:2017b, Zhou:2018, Ratcliffe:2018}. However, these nonclassical properties also considerably modify the kinds of control strategies and mechanisms available to quantum systems.  

One key example of the differences is the role of measurement. It is a given in classical control that one can measure the system and act upon it based on the information extracted about the system. However, for a quantum system measurement itself changes the state of the system and this has to be carefully accounted for in the design of many closed-loop control protocols~\cite{Wiseman:1994, Wiseman:1996, Doherty:2000,Handel:2005, Szigeti:2009,Hamerly:2013, Hush:2013}. Although measurement backaction is usually considered undesirable---an unwanted effect to be minimized---from another perspective measurement is an extra ``control knob'' unavailable in the classical context, which can be used to develop new control strategies for quantum dynamical systems \cite{Blok:2014, Szigeti:2014}. In particular, adaptive measurements have been used to improve phase estimation~\cite{Wiseman:1995}, in quantum state preparation~\cite{Ralph:2005}, and to enhance the precision of quantum measurements~\cite{Higgins:2007}. 

In this paper, we explore how this uniquely quantum knob can be used to control the dynamics of a chaotic system. Classically, controlling these systems is both a significant and nontrivial problem. In some situations it is desirable to induce chaotic dynamics, as in the case of embedding data into chaotic signals for secure transmission of information~\cite{Lau:2003}. However, in other cases the task is to lock the system to stable orbits, as when aiming to regularize the behavior of cardiac rhythms~\cite{Ferreira:2014} or improve energy harvesting in cantilever devices~\cite{Erturk:2011, Kumar:2016}. In many of these stabilization problems, feedback methods are used to turn an originally unstable orbit embedded in the chaotic attractor into a regular one~\cite{Ott:1990, Pyragas:1992}. In this work, we show that transitioning at will from chaos to regularity is possible by using a real-time adaptive measurement protocol. In particular, our protocol combines the tunability of quantum measurement backaction on the quantum state with the underlying geometry of the classical dynamical system. This opens up regimes of control not available to open-loop control schemes.

This quantum control strategy cannot be borrowed straightforwardly from an analogous classical problem, not only because of the aforementioned peculiarities of quantum measurement, but also due to subtleties associated with identifying emergent quantum chaotic orbits. In a closed quantum system, coherent interference effects cause a breakdown in the correspondence principle such that chaotic classical dynamics do not emerge when the underlying quantum model is taken to the macroscopic limit~\cite{Ehrenfest:1927}. However, in \emph{open} quantum systems, decoherence destroys such quantum interference effects~\cite{Zurek:1994}, allowing emergent chaotic dynamics in the classical limit~\cite{Ott:1984,Dittrich:1987, Zurek:1994, Spiller:1994, Habib:1998, Pattanayak:2003, Carvalho:2004b, Carvalho:2004c}. In particular, by considering stochastic unravelings of an open quantum system, which are physically associated with making particular continuous measurements on the system~\cite{Wiseman:2001b, Wiseman:2010, Rigo:1996}, we can observe chaos in the conditional system dynamics~\cite{Rigo:1996,Spiller:1994}. The stochastic unravelings  allow chaos to be identified and quantified with the quantum Lyapunov exponent~\cite{Bhattacharya:2000,Ota:2005,Habib:2006,Kapulkin:2008,Pokharel:2018} and also provide the necessary ingredient for a closed-loop feedback control scheme. 

Previously, we showed that the behavior of the system can be chaotic or not depending on the initial (and fixed) choice of measurement, due to the interplay between the interference effects induced by the nonlinear dynamics and the effectiveness of the measurement in destroying them~\cite{Eastman:2017}. This sensitivity to measurement choice was shown to be absent both in the macroscopic limit, where the effects of quantum measurement are naturally expected to disappear, and in a highly-quantum regime, where noise dominates and measurement choice becomes irrelevant. Although the system behaves chaotically in the former case, as in the classical analog, in the latter, chaos is suppressed by quantum effects. As the main outcome of the control protocol presented here, we are able to show that a judicious real-time choice of measurement can induce chaotic behavior deeper in the quantum regime, effectively pushing the quantum-classical boundary further towards the microscopic domain.

\section{Quantum Duffing Oscillator}
To illustrate our adaptive protocol, we consider a driven-damped Duffing oscillator~\cite{Duffing:1918}, a model that has been extensively used in the investigation of chaotic dynamics in open quantum systems~\cite{Eastman:2017,Pokharel:2018, Ota:2005,Ralph:2017a,Brun:1996,Schack:1995}. 
The model consists of a particle that oscillates in a double-well potential that is periodically tilted by an external driving force with amplitude $g$ and frequency $\Omega$. The dimensionless quantum Hamiltonian describing this model is given by  
\begin{equation}
\label{eq:hamiltonian}
\hat{H} = \frac{1}{2} \hat{P}^2 + \frac{\beta^2}{4}\hat{Q}^4 - \frac{1}{2} \hat{Q}^2 +  \frac{\Gamma}{2}(\hat{Q}\hat{P}+ \hat{P}\hat{Q}) -\frac{g}{\beta}\hat{Q} \cos{(\Omega t)},
\end{equation} 
where time is in units of the trap period $2\pi/\omega_0$ and $\hat{Q}=\hat x / \sqrt{\hbar / (m \omega_0)} $ and $\hat{P}= \hat p / \sqrt{\hbar m \omega_0}$ are, respectively, the dimensionless position and momentum operators for a single particle of mass $m$. The first term in the Hamiltonian describes the kinetic energy, the quartic and quadratic terms in $\hat{Q}$ describe the double-well potential, and the last term describes the periodic driving of the particle. The dimensionless parameter $\beta^2 = \hbar / (ml^2 \omega_0)$ defines the scale of the phase space relative to Planck's constant~\cite{Brun:1996,Ota:2005,Kapulkin:2008} (where $l$ characterizes the size of the system). 
%\jke{Tuning this parameter effectively scales the size of the dynamics with respect to the quantum noise which enables scanning of the transition from quantum regime to approaching the classical regime ($\beta \to 0$).} 
A larger $\beta$ is therefore associated with a regime where quantum fluctuations have a larger effect on the oscillator dynamics. Thus, by tuning $\beta$ we can study the transition from the quantum regime to the classical regime ($\beta \to 0$).

To include damping, we model the quantum dynamics through the master equation 
\begin{equation}
\label{eq:ME}
\dot{\rho} = -i[\hat H,\rho] + \left(\hat L \rho \hat L^\dagger - \frac{1}{2}\{\hat L^\dagger \hat L, \rho \}\right),
\end{equation}
where dissipation effects arise from choosing the system-environment coupling, $\hat L = \sqrt{\Gamma} (\hat{Q} + i\hat{P}) = \sqrt{2 \Gamma} \hat a$, to be proportional to the annihilation operator of the harmonic oscillator. 

In the classical limit ($\beta \to 0$), we can make the identifications $\langle \hat{Q} \rangle \rightarrow x_{cl}$ and $\langle \hat{P}\rangle\rightarrow p_{cl}$ such that the equations of motion for $\langle \hat{Q} \rangle$ and $\langle \hat{P} \rangle$ correspond to the dimensionless classical dynamics given by~\cite{Brun:1996,Ota:2005,Kapulkin:2008,Eastman:2017}
\begin{equation}
\ddot{x}_{cl} + 2\Gamma\dot{x}_{cl} + \beta^2{x}^3_{cl} - x_{cl} = \frac{g}{\beta}\cos{(\Omega t)}.
\label{eq:class}
\end{equation}
Although the scaling factor $\beta$ is crucial in determining the role of quantum effects in the dynamics, classically it is a trivial scaling factor due to the definition of $x_{cl}$ and $p_{cl}$. Indeed, for rescaling $X\equiv \beta x_{cl}$, the classical equation of motion is independent of $\beta$. Note also that the quantum dissipation, given in terms of $\hat L$, is symmetric with respect to position and momentum. The extra term proportional to the damping rate $\Gamma$ in the Hamiltonian~(\ref{eq:hamiltonian}), breaks this symmetry in such a way that the dissipative force is proportional to the velocity, exactly as expected in the classical limit.

\begin{figure}[ht]
\centerline{\includegraphics[width=0.95\linewidth]{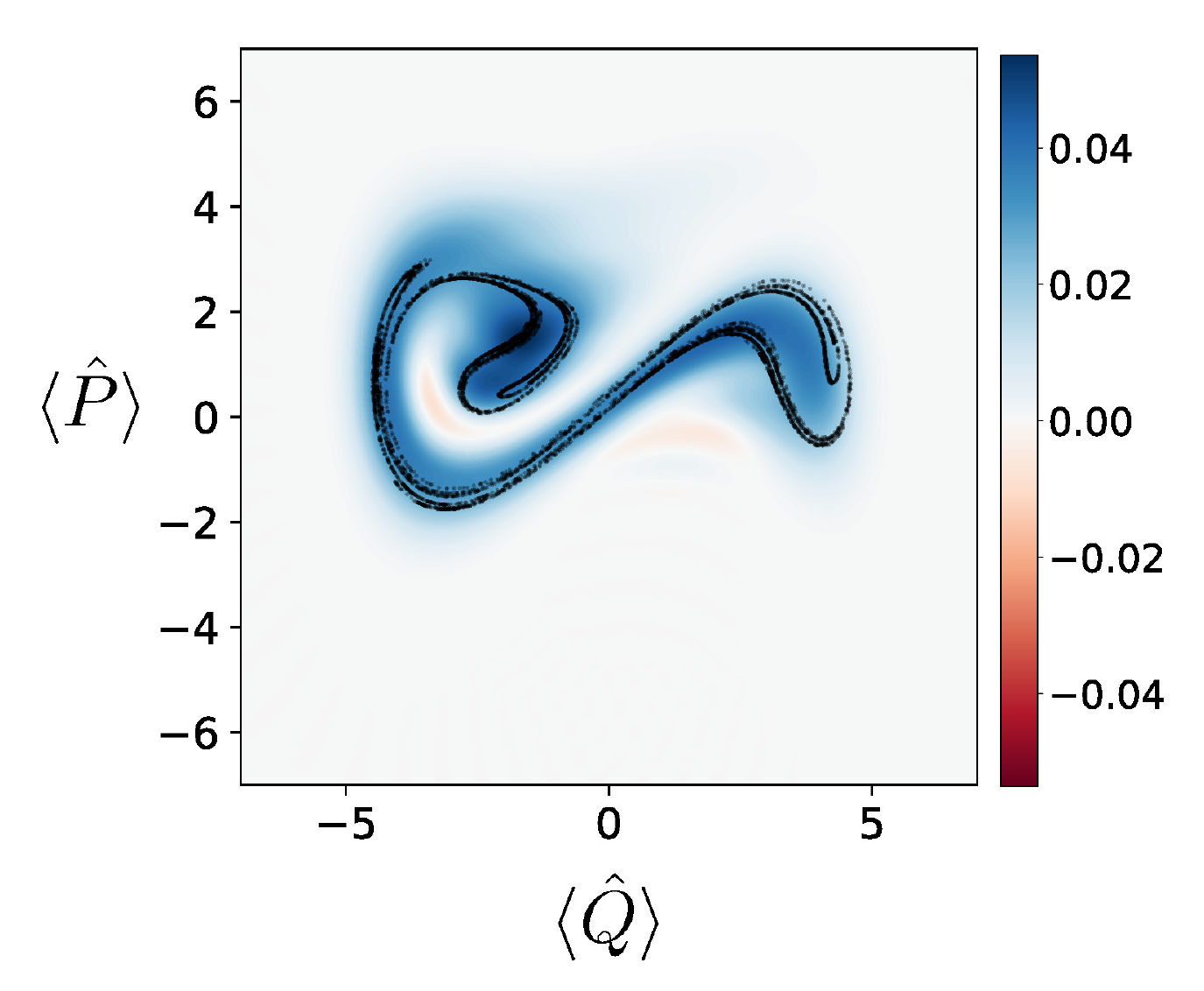}}
\caption{Wigner function for the steady state of the unconditional dynamics given by the master equation~(\ref{eq:ME}), for the dimensionless parameters $\Gamma = 0.10$, $g=0.3$, and $\Omega = 1$.  The Poincar\'e section of the classical Duffing oscillator is also overlaid for these parameters (black dots). The system exhibits chaos for these parameters, as seen by the emergence of the strange attractor and the positive Lyapunov exponent $\lambda_{cl} = 0.16$. Here the Wigner function for the unconditional state also follows the shape of the strange attractor in phase space, which is a signature of chaotic dynamics. The scaling parameter $\beta=0.3$ was chosen to allow for a direct comparison with the results in Figs.~\ref{fig:projection} and ~\ref{fig:wig}. }
%\caption{\jke{ Change Caption} Poincar\'e section of the classical Duffing oscillator for the dimensionless parameters $\Gamma = 0.10$, $g=0.3$, $\Omega = 1$. For this choice of parameters, the system exhibits chaos, as seen by the emergence of the strange attractor and the positive Lyapunov exponent $\lambda_{cl} = 0.16$. The scaling parameter $\beta=0.3$ was chosen to allow for a direct comparison with the quantum results in Fig.~\ref{fig:projection} and Fig.~\ref{fig:wig}.}
\label{fig:poincare}
\end{figure}

Depending on the parameters, the classical model described by Eq.~(\ref{eq:class}) exhibits chaotic dynamics as illustrated by the strange attractor in phase space shown by the black dots in Fig.~\ref{fig:poincare}. 
The steady state of the Wigner function, obtained by numerically solving Eq.~(\ref{eq:ME}), is also shown in Fig.~\ref{fig:poincare} for the same set of parameters. This illustrates that the Wigner function of the ensemble-averaged quantum state broadly matches the strange attractor, which is a signature of chaotic dynamics. However, the degree of chaos cannot be quantified via the unconditional dynamics of Eq.~(\ref{eq:ME}), since any two initial states evolve to the same asymptotic state, giving a negative Lyapunov exponent. This does \emph{not} mean that chaos is not present; indeed, the same problem would arise in classical chaos if one decided to calculate classical Lyapunov exponents by using the separation of average trajectories over a classical ensemble, rather than the separation of two classical trajectories. To define the degree of chaos via a \emph{quantum} Lyapunov exponent, we need to use a conditional quantum trajectory approach that has a direct comparison with the classical trajectory approach~\cite{Spiller:1994, Schack:1995, Bhattacharya:2003, Ghose:2004}. 

\section{Continuous measurement of an open quantum system} \label{sec_cont_meas}
The master equation~(\ref{eq:ME}), describes the ensemble-averaged evolution of the open quantum system. However, implementing a closed-loop control scheme that depends on the monitored real-time dynamics requires a description of a single experimental realization (or trajectory). This is provided by stochastic unravelings of the master equation, which correspond to the evolution of the quantum state conditioned on a continuous measurement record~\cite{Carmichael:1993,Percival:1998a, Wiseman:2001b,Rigo:1997}.

Here we consider the class of diffusive quantum trajectories which, in its most general form, is described by the Ito stochastic Schr\"odinger equation (SSE)~\cite{Rigo:1997,Wiseman:2001b}:
\begin{eqnarray}
\label{eq:SSEITO}
\mathrm{d} \ket{\psi} &=& \left(-i \hat H  -\frac{{\hat L}^\dagger \hat L}{2} + \langle {\hat L}^\dagger \rangle \hat L -\frac{ \langle {\hat L}^\dagger \rangle \langle \hat L \rangle}{2}  \right) \ket{\psi} \mathrm{d}t \nonumber \\  &+& \left(\hat L - \langle \hat L \rangle \right)\ket{\psi}  \mathrm{d}\xi,
\end{eqnarray}
where the noise term $\mathrm{d}\xi$ is a complex Wiener process with zero mean ($\mathbb{E}[d\xi]=0$) and correlations
\begin{equation}
\label{eq:corr}
\mathrm{d}\xi \, \mathrm{d}\xi^* = \mathrm{dt} \qquad \mathrm{and} \qquad \mathrm{d}\xi \, \mathrm{d}\xi = u\, \mathrm{dt}, 
\end{equation}
with $u$ being a complex number satisfying $\vert u\vert \le 1$~\cite{Rigo:1997,Wiseman:2001b}. In what follows, we choose $u= \exp{(-2i\phi)}$ so that $\mathrm{d}\xi = \exp{(-i\phi)}~\mathrm{d}W$, where $dW$ is a real noise of zero mean and $dW^2 = dt$. Physically, this choice corresponds to a continuous measurement of the quadrature operator $\hat{X}_\phi = [\exp{(-i\phi)} \hat{a} + \exp{(i\phi)} \hat{a}^\dagger]/\sqrt{2}$. 
Experimentally, this could be achieved by performing a standard balanced homodyne detection on the output of the system, as shown in Fig.~\ref{fig:adapscheme}. The output channel $\hat{L} = \sqrt{2 \Gamma} \hat{a}$ is combined with a local oscillator (LO) of phase $\phi$ at a beam splitter, while the readings at the detectors are subtracted to yield a measurement signal $I \mathrm{d}t =\sqrt{\Gamma} \langle \hat{X}_\phi \rangle + \mathrm{d}W$~\cite{Wiseman:2001b}. The phase $\phi$ of the LO is a controllable parameter that determines the quadrature to be measured. For instance, $\phi = 0$ results in a measurement of $\hat{Q} = \hat{X}_{\phi=0}$, whereas $\phi=\pi/2$ gives a measurement of $\hat{P} = \hat{X}_{\phi=\pi/2}$.

Within the context of quantum chaos, this quantum trajectory approach has proven useful in the investigation of the quantum-classical transition~\cite{Spiller:1994,Brun:1996,Rigo:1996,Rigo:1997,Ghose:2003}. Furthermore, it offers a way to calculate quantum Lyapunov exponents, thereby unambiguously quantifying the degree of chaos within the system~\cite{Bhattacharya:2000,Ota:2005,Habib:2006,Kapulkin:2008,Eastman:2017,Pokharel:2018}. Similar to the classical protocol~\cite{Wolf:1985}, this is done by following the separation of two initially close wave-packet centroids in phase space $(\langle \hat{Q}\rangle,\langle \hat{P}\rangle)$ evolving according to Eq.~(\ref{eq:SSEITO}) under the same noise realization~\cite{Eastman:2017,Pokharel:2018}. 

Specifically, the quantum Lyapunov exponent is defined as 
\begin{equation}
	\lambda =\lim_{t \to \infty} \lim_{d_0 \to 0} \frac{\ln{\left(d_t/d_0\right)}}{t},
\end{equation}
where $d_t=[\Delta Q(t)^2 + \Delta P(t)^2]^{1/2}$ is the dimensionless phase-space distance between two quantum trajectories with differences in the average position and average momentum of the two trajectories given by $\Delta Q(t) = \langle \hat{Q}_{1} \rangle - \langle \hat{Q}_{2} \rangle$ and $\Delta P(t) = \langle \hat{P}_{1} \rangle - \langle \hat{P}_{2} \rangle$, respectively. The two quantum trajectories are initially prepared in coherent states displaced (in phase space) from each other by a small distance $d_0 = d_{t=0}$ (i.e., $|\alpha_1 \rangle = |\alpha\rangle$ and $|\alpha_2 \rangle = |\alpha + d_0\rangle$), and then evolved stochastically via Eq.~(\ref{eq:SSEITO}) under the same noise realization, which corresponds to the same measurement record. Using this approach, it was shown in Ref.~\cite{Eastman:2017} that the choice of measurement angle $\phi$ has a direct effect on the quantum Lyapunov exponent and, therefore, on the emergence of chaos in quantum systems.

\section{Adaptive measurement protocol for controlling chaos}
\label{sec:adaptive}
The continuous measurement approach described in Sec.~\ref{sec_cont_meas} naturally sets the scene for our main result: the design of a protocol to control chaos by using a tunable, and experimentally accessible, parameter. The parameter in question, the LO phase $\phi$, is intrinsically linked to the measurement backaction, making our control mechanism fundamentally quantum in nature. 

\begin{figure}[ht]
\centerline{\includegraphics[width=\linewidth]{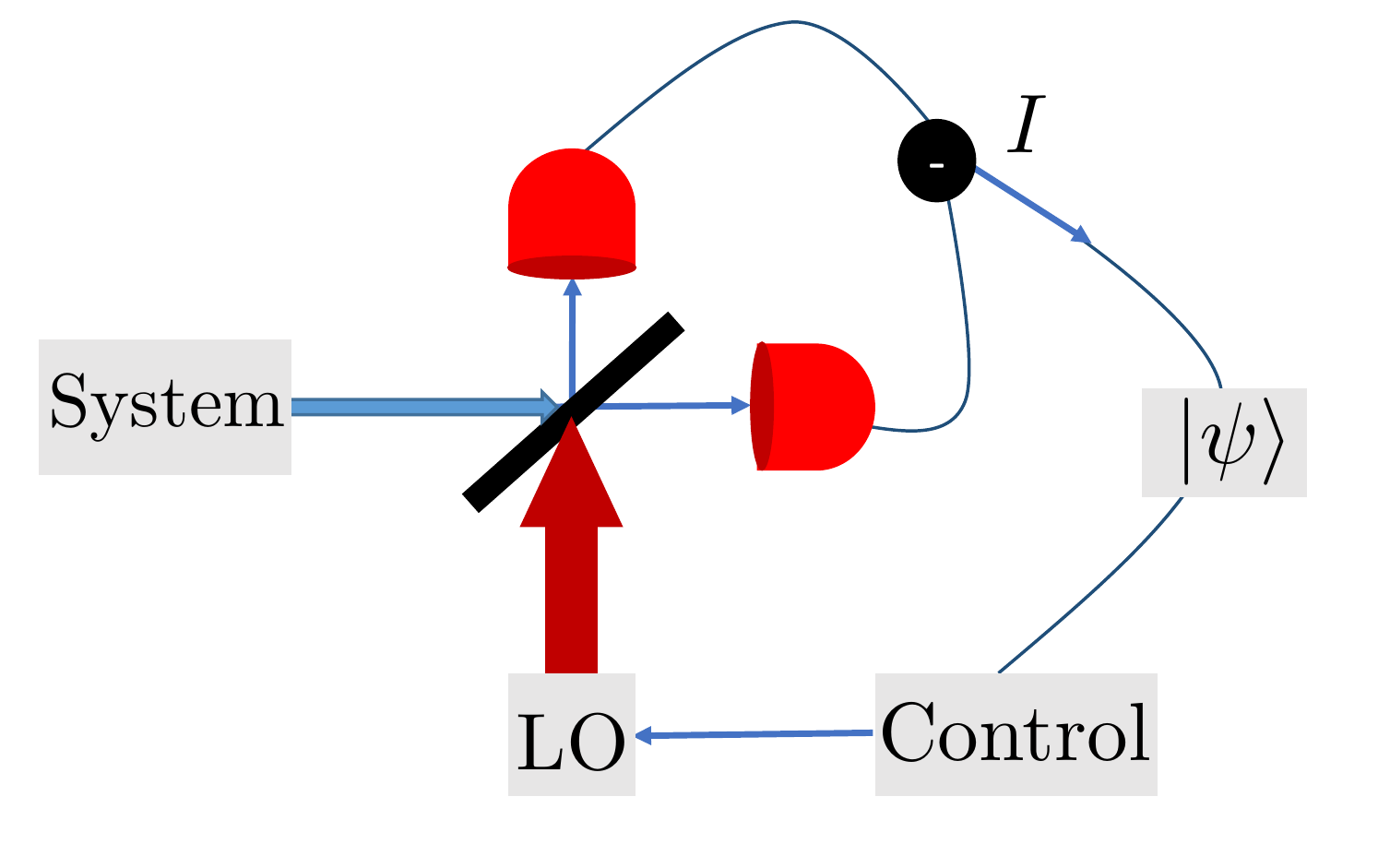}}
\caption{Adaptive measurement scheme in a quantum optics setup. The state-dependent controller chooses the LO phase $\phi$ at each time step in order to change the measurement backaction applied to the system, which changes the evolution as desired.}
\label{fig:adapscheme}
\end{figure}

The scheme we consider is shown in Fig.~\ref{fig:adapscheme}. The continuous monitoring of the system gives a measurement signal, $I(t)$, that allows for a real-time estimate of the quantum state. In possession of this information, one can then design a feedback action to influence the system. Motivated by the effect that measurement has on the system dynamics~\cite{Eastman:2017}, here we propose to adaptively change the phase $\phi$ in real time, with the intent to control the Lyapunov exponent of the system.

\begin{figure*}[ht]
\centerline{\includegraphics[width=\textwidth]{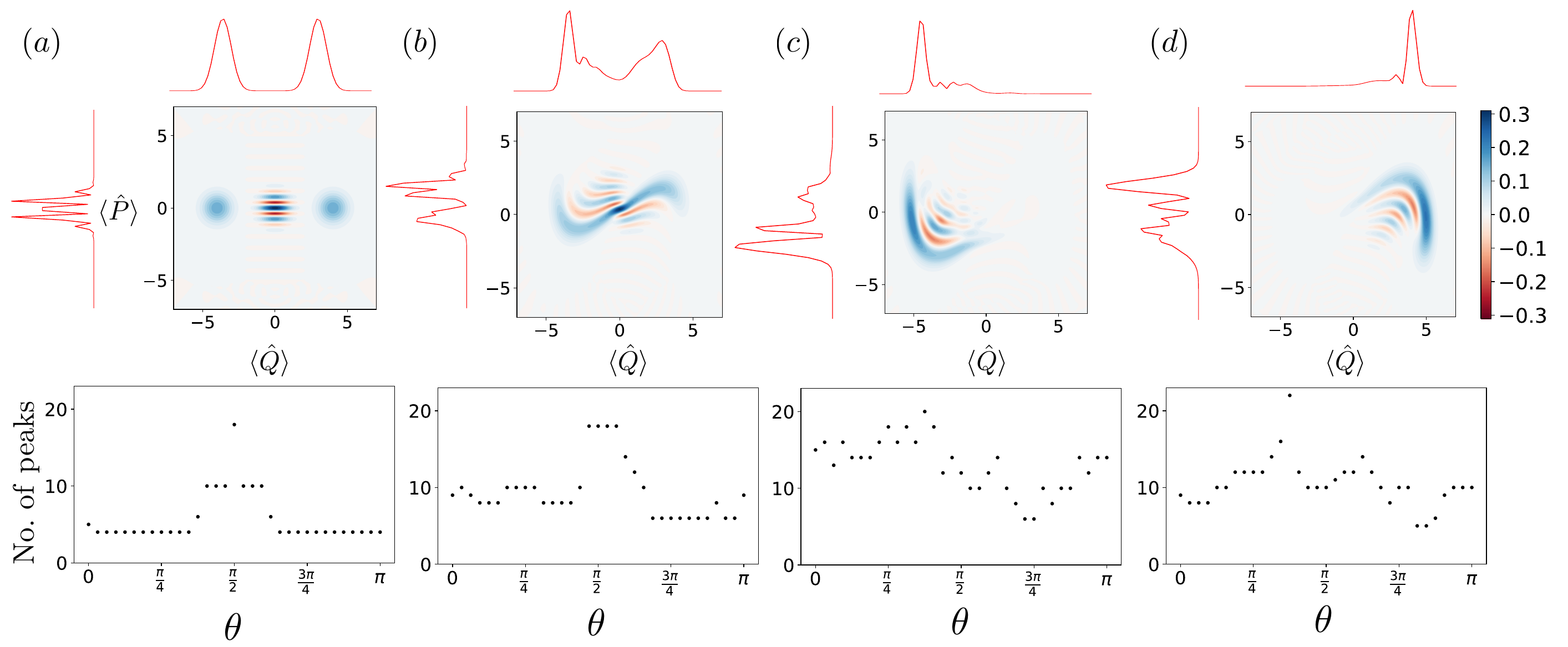}}
\caption{Wigner functions and corresponding phase quadrature projections $X_{\theta = 0}$ and $X_{\theta = \pi/2}$ for (a) a Schr\"odinger cat state $|\psi_{cat}\rangle \propto |\alpha \rangle + |-\alpha \rangle$ and (b)-(d) three snap shots typically seen in the evolution of the quantum Duffing oscillator. Here $\theta = 0$ (top) and $\theta = \pi/2$ (side) are the only projections plotted. The number of peaks in the probability distributions is plotted as a function of quadrature $\theta$ for 32 different angles. The maximum in the number of peaks corresponds to the direction perpendicular to the interference fringes ($\theta_\text{max}-\theta_f=\pi/2$). Note that the number of peaks in the bottom plots does not equal the number of peaks seen in the probability distributions. This is a numerical noise associated with counting every turning point and does not affect the outcome of the search.}
\label{fig:projection}
\end{figure*}

The design of an effective control strategy relies on first understanding how the feedback action affects the system. For that, we recall a fact observed in Ref.~\cite{Eastman:2017}: The stretches and foldings induced by the chaotic dynamics generate interference fringes in the Wigner function of the system (see top panel of Fig.~\ref{fig:projection}), and these lead to the suppression of chaos in the quantum regime.  Since these interference fringes are associated with quantum coherence, destroying them shifts the dynamics towards the classical chaotic behavior. Therefore, in order to enhance (suppress) chaos, our state-dependent controller chooses the LO phase $\phi$ such that the measurement destroys the interference fringes in the state's Wigner function at the fastest (slowest) possible rate. More precisely, this rate of fringe destruction is determined by the direction of the interference fringes in phase space ($\theta_f$) relative to the axis of measurement (determined solely by $\phi$), with fast destruction rates occurring when these axes are aligned. Our control protocol then consists of estimating the fringe structure in real time and picking a $\phi(t)$ that would maximize the control objective.  

Automating the process of determining the direction of the interference fringes in the Wigner function can be done by examining the probability distributions for different quadrature measurements:
\begin{eqnarray}
\label{eq:prob}
P_{X_\theta}&=& \vert\langle X_\theta | \psi\rangle\vert^2, 
\end{eqnarray}
where $\ket{X_\theta}$ is an eigenstate of the quadrature operator $\hat{X}_\theta$. To understand how this can be used to estimate the fringe structure, let us look at the particular case of the Schr\"odinger cat state $|\psi_{cat}\rangle \propto |\alpha\rangle + |-\alpha\rangle$ shown in Fig.~\ref{fig:projection}(a). Projection onto the $\hat{X}_{0}$ quadrature is given by the top red plot in Fig.~\ref{fig:projection}(a). Here, a measurement of $\hat{X}_0$ distinguishes between the two coherent states, resulting in two peaks. In contrast, the projection onto the $\hat{X}_{\pi/2}$ quadrature (the red plot to the left of the Wigner function plot) reveals the overlap of the two coherent states, resulting in interference fringes and a large number of peaks. As shown directly below the Wigner function plot, looking at the number of peaks as a function of projection angle $\theta$ reveals that the peak distribution is narrowly centered around $\theta = \pi/2$ [the $\langle \hat{P} \rangle$ axis], which is perpendicular to the interference fringe axis. This shows that the angle that maximizes the number of peaks ($\theta_\text{max}$) is a good indicator of the direction that is perpendicular to the fringes in the Wigner function. 

In the actual quantum Duffing oscillator, the nonlinear dynamics lead to interference fringe patterns with more complexity than those of a Schr\"odinger cat state. Examples of the Wigner functions for typical evolved states that arise during this evolution are plotted in Figs.~\ref{fig:projection}(b)-\ref{fig:projection}(d). Although more complicated, these Wigner functions still present a reasonably-well-defined direction in the fringe structure, which can be determined by finding the angle that leads to the maximum number of peaks in $P_{X_\theta}$, as explained above. 

In summary, our protocol consists of the following steps:
\begin{enumerate}[(i)]
\item{Starting from a given $\ket{\psi (t)}$, calculate $P_{X_\theta}$ for various $\theta$;}
\item{Count the number of peaks for each $P_{X_\theta}$ and find $\theta_\text{max}$;}
\item{To maximize (minimize) the Lyapunov exponent, choose $\phi(t)=\theta_{f}=\theta_\text{max}-\pi/2$ ($\phi(t)=\theta_\text{max}$);}
\item{Use the value of $\phi(t)$ from (iii) in Eq.~(\ref{eq:SSEITO}) to calculate the new state $\ket{\psi (t+dt)}$;}
\item{Repeat steps (i) to (iv).}
\end{enumerate}
Full details of the numerical implementation of these steps are given in the appendix.

\section{Results}
\begin{figure}[ht]
\centerline{\includegraphics[width=0.95\linewidth]{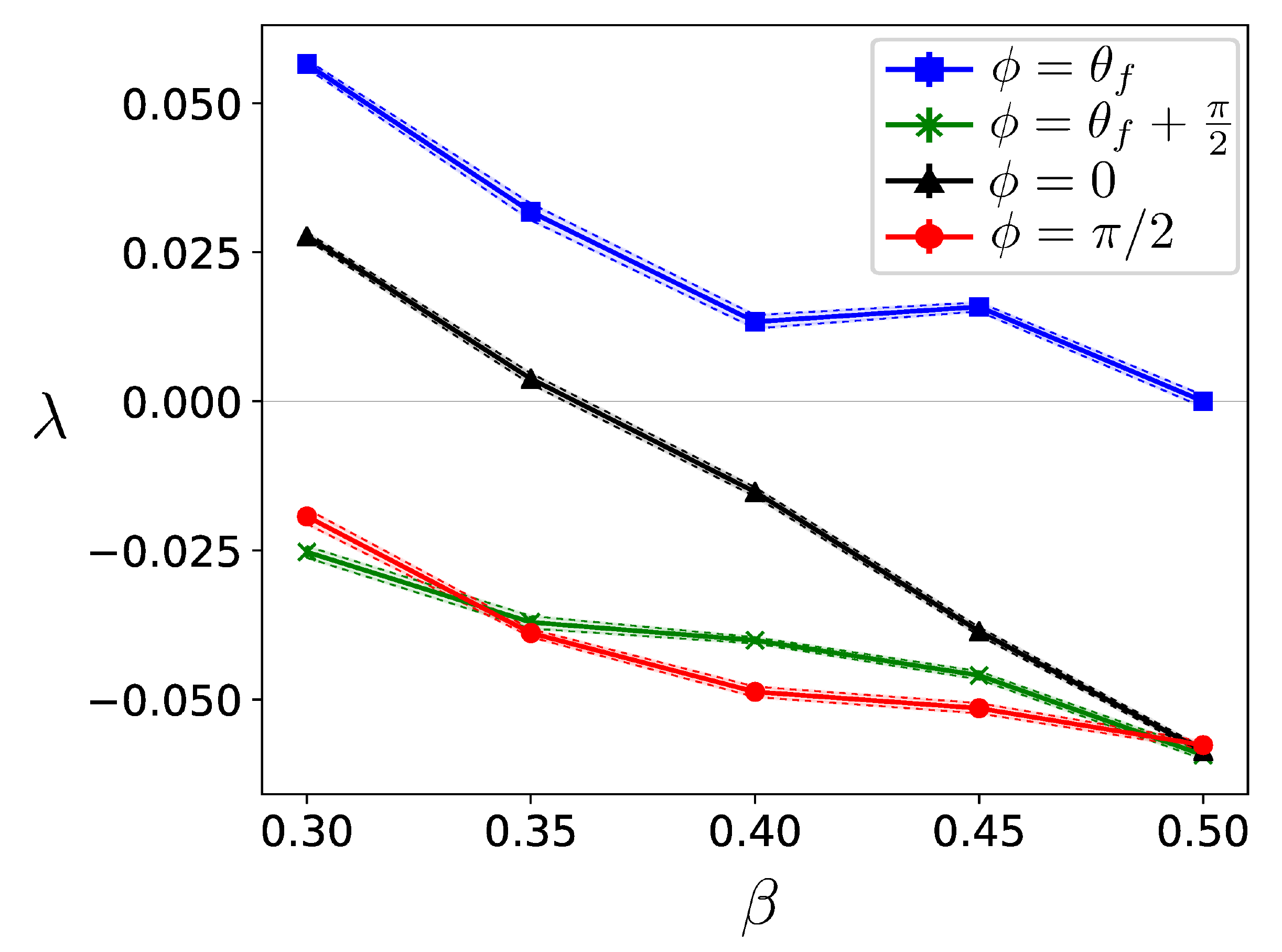}}
\caption{The quantum Lyapunov exponent ($\lambda$) as a function of $\beta$ for adaptive measurements ($\phi=\theta_f$, blue squares, and $\phi = \theta_f + \pi/2$, green crosses) and fixed LO measurements ($\phi = 0$, black triangles, and $\phi = \pi/2$, red circles).  Here $\Gamma = 0.10$, $g = 0.3$ and $\Omega = 1$, and the classical system is chaotic with $\lambda_{cl}=0.16$. Each point is averaged over 10 different noise realizations and the shaded area within the dashed lines signifies twice the standard error.
}
\label{fig:allbeta}
\end{figure}

We implemented the adaptive measurement scheme described in Sec.~\ref{sec:adaptive} for a range of scaling parameters $\beta$ (spanning the transition from the quantum regime to the classical regime) and two distinguishable strategies: maximization and minimization of the Lyapunov exponent ($\lambda$). The results are shown in Fig.~\ref{fig:allbeta} for both cases, specifically, where the LO phase is set to always measure along an axis parallel ($\phi= \theta_f$, blue line, square points) or perpendicular ($\phi=\theta_f+\pi/2$, green line, crosses) to the interference fringes. To assess the effectiveness of our adaptive protocol, we compare with the best nonadaptive strategy by displaying the curves that maximize (black line, triangles) and minimize (red line, circles) $\lambda$ for a \emph{fixed} LO phase. 

\begin{figure*}[ht]
\centerline{\includegraphics[width=\textwidth]{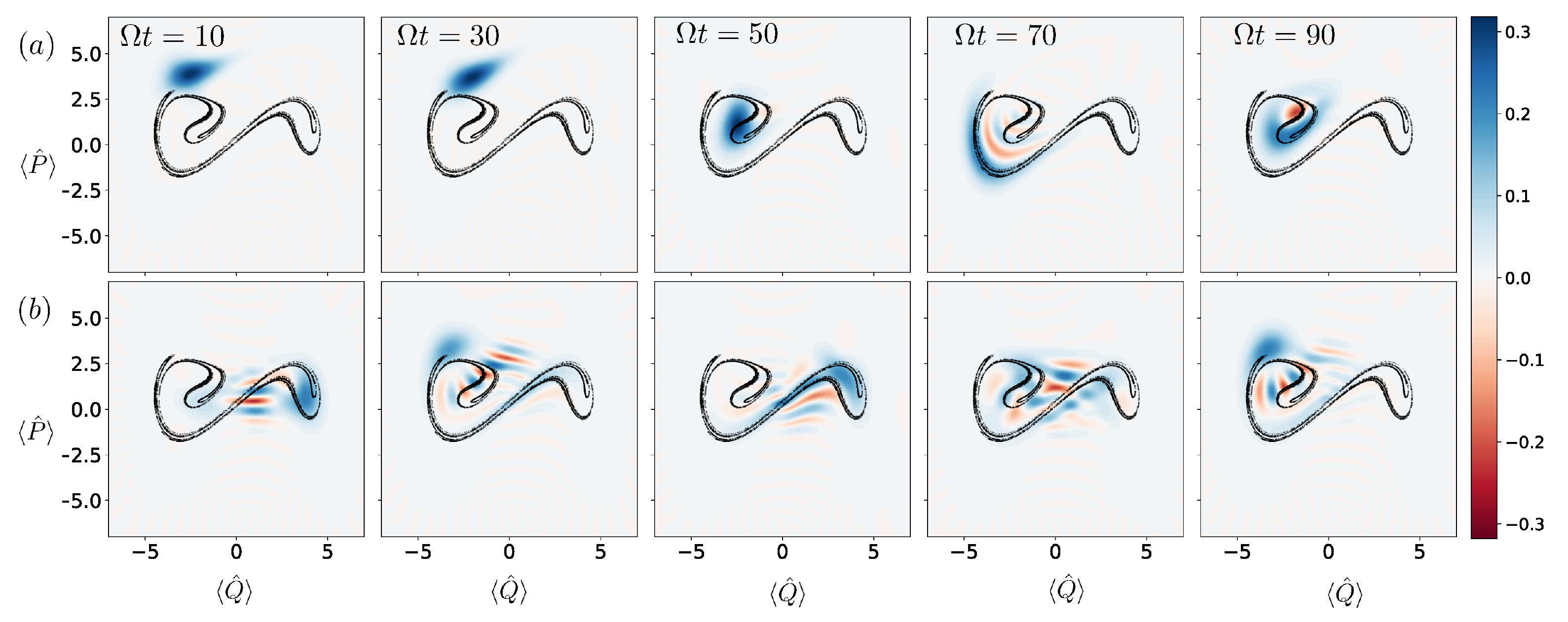}}
\caption{Snap shots of the Wigner function for the first 100 cycles of the driving for both adaptive measurements [(a) $\phi=\theta$, and (b) $\phi=\theta + \pi/2$]. The snap shots only show a single quantum trajectory (noise realization); however, all trajectories have similar evolution to that depicted here. The corresponding Lyapunov exponents are (a) $\lambda = 0.057 \pm 0.001$ and (b) $\lambda =  -0.025 \pm 0.001$.}
\label{fig:wig}
\end{figure*}

The adaptive maximization strategy leads to Lyapunov exponents that are always larger than the best fixed-angle scenario ($\phi=0$). By destroying coherent interference effects and localising the state faster, the adaptive case allows the quantum system to track the classical chaotic dynamics more closely, increasing $\lambda$. Further evidence of this is provided by looking at the dynamical evolution of the Wigner function (see Fig.~\ref{fig:wig}, top), showing states that are more localized and possess less interference, and are therefore more classical in nature. The opposite adaptive strategy, the one designed to suppress chaos, also works effectively, giving negative Lyapunov exponents for all values of $\beta$. In this case, the adaptive choice of monitoring angle leads to the preservation of quantum interference effects and therefore to highly nonclassical states with a large spread in phase space, as seen in the Wigner functions of Fig.~\ref{fig:wig} (bottom). 

Interestingly, the adaptive $\lambda$-maximization scheme gives positive Lyapunov exponents for much larger values of $\beta$ (up to $0.5$), showing that the adaptive protocol pushes the emergence of chaos deep into the quantum regime---and much further than what is possible with a fixed LO phase. 
%This is a remarkable behavior given that, at this scale, quantum uncertainties should be overshadowing the underlying chaotic dynamics, making the choice of measurement direction, adaptive or not, irrelevant. This is evident in Fig.\ref{fig:allbeta}, where the Lyapunov exponent for all monitoring schemes other than $\lambda$-maximization converge to roughly the same value. Yet, in stark contrast, our maximization protocol is able to sustain chaotic dynamics even at this scale. 
This is remarkable behavior given that quantum noise is expected to dominate the dynamics at these large values of $\beta$, and so one would think that the choice of measurement is irrelevant. This is clearly the case for the fixed measurement (see Fig.~\ref{fig:allbeta}), where the quantum Lyapunov exponent for all monitoring schemes other than $\lambda$-maximization converge to roughly the same negative value, indicating regular dynamics. In stark contrast, our $\lambda$-maximization protocol is able to sustain chaotic dynamics even at this scale. %This suggests that there is still some small dependence on the measurement choice and that coherences will still play a role in suppressing chaos in this regime.

%\jke{Although our adaptive protocol can significantly enhance chaos, the enhancement of regularity using the adaptive $\lambda$-minimisation scheme provides little benefit over the fixed measurement. This is a downfall of the metric that we have chosen to use (Eq.~\ref{eq:prob}). As can be clearly seen from  Fig.~\ref{fig:wig}(b), in particular at time $\Omega t = 70$, the state is highly non classical and delocalized. As a result of this the direction of the interference fringes is not well defined and this is clearly an issue for our metric which relies on the direction of fringes to inform the next measurement choice. For our $\lambda$-maximisation this isn't an issue, since our state is much more localised. But for the case of $\lambda$-minimisation, the highly delocalised regime is encouraged, which in turn makes it harder to choose the correct angle for the measurement. }

Although our adaptive $\lambda$-maximization scheme can significantly enhance chaos, the adaptive $\lambda$-minimization scheme does not provide significantly enhanced regularity over the fixed measurement. This is a consequence of using metric~(\ref{eq:prob}) to choose the measurement quadrature angle $\phi$ at each time point. The aim is to find the direction of interference fringes in the Wigner function, and choose a measurement angle parallel (perpendicular) to this direction in order to enhance (suppress) chaos. However, the metric~(\ref{eq:prob}) becomes less effective when the state is highly nonclassical and delocalized. This is shown clearly in the Wigner function plots of Fig.~\ref{fig:wig}(b), in particular at time $\Omega t = 70$. In this case, the large degree of delocalization means that there is no well-defined single direction of interference fringes. Consequently, in this regime the adaptive control does not provide a substantially improved performance over a fixed-angle measurement.
When trying to suppress chaos by picking a measurement that has the least deleterious effect on quantum interferences, it is exactly this highly delocalized regime that is encouraged. Therefore, it is unsurprising that our adaptive measurement protocol provides little benefit over a fixed measurement angle, if the goal is to suppress chaos. In contrast, our metric is more effective when the Wigner function is localized and the fringe direction better defined [see Fig.~\ref{fig:wig}(a) for $\Omega t = 70$]. This is the scenario arising from our strategy to enhance chaos: choosing measurements that destroy coherence and keep the state localized.

\section{Discussion}
We briefly discuss the experimental prospects of realizing both the driven-damped quantum Duffing oscillator and our adaptive measurement protocol. Superconducting circuits are excellent candidate systems, due to their flexible architecture, wide range of experimental parameters, and the existence of demonstrated continuous probing~\cite{Wendin:2017}. Specifically, superconducting circuits in a parallel circuit configuration (i.e., a rf-SQUID) could be used to experimentally realize a quantum Duffing oscillator~\cite{Manucharyan:2007,Ralph:2017a}. For the scheme proposed in Ref.~\cite{Ralph:2017a}, $\beta^2 = e^2 / [3 \hbar \omega C_p(1-L_p/L_J)]$, where $\omega = 1 / \sqrt{C_p L_p}$, $C_p$ is the capacitance of the Josephson junction in the circuit, $L_p^{-1} = L_J^{-1} - L_p^{-1}$ is the parallel inductance formed from the Josephson inductance $L_J$ and the geometric inductance $L_{pe}$, and $e$ is the charge of an electron. Using typical experimental parameters from Ref.~\cite{Boutin:2017}, we estimate that $\beta \sim 0.4$ is currently achievable which, as shown in Fig.~\ref{fig:allbeta}, is a regime ideally suited for observing measurement-dependent effects on the emergence of chaos.

Realizing our scheme with ultracold atomic gases is another potential option. Ultracold atomic experiments have previously been used to experimentally investigate the emergence of chaos in the quantum kicked rotor~\cite{Raizen:1996aa, Duffy:2004, Tomkovic:2017}. A Bose-Einstein condensate (BEC) provides the high optical densities needed for real-time nondestructive imaging~\cite{Wigley:2016,Everitt:2017}. A noninteracting BEC gives the single-particle behavior required to realize the driven-damped quantum Duffing oscillator. A noninteracting gas can be achieved by using an extremely dilute sample or by extinguishing the interactions via a Feshbach resonance~\cite{McDonald:2014,Everitt:2017}. The required double-well potential could be created by superimposing a Gaussian barrier on a harmonic potential: 
\begin{align}
	\hat{V}_\text{exp}	&= \frac{1}{2} m \omega_0^2 \hat{x}^2 + A e^{-\hat{x}^2/2\sigma^2} \notag \\
    					&\approx \hbar \omega_0 \left[ \frac{1}{2} \left( 1 - \frac{A}{m \omega_0^2 \sigma^2}\right) \hat{Q}^2 + \frac{1}{4} \left( \frac{\hbar A}{2 m^2 \omega_0^3 \sigma^4}\right) \hat{Q}^4\right].
\end{align}
The choice of barrier height $A = 2 m \omega_0^2 \sigma^2$ realizes the needed potential [see Eq.~(\ref{eq:hamiltonian})] with $\beta^2 = \hbar/ (m\omega_0 \sigma^2)$. There are a number of techniques for creating this potential, including via an optical lattice~\cite{Spagnolli:2017} or spatial light modulation~\cite{Gauthier:2016}. For the 780 nm transition of $^{85}$Rb, a barrier waist of $\sigma \sim 10 \mu$m is easily achievable. For typical trapping frequencies $\omega_0 \in 2\pi \times [5,100]$ Hz, this gives $\beta \sim 0.1-0.5$.

These simple estimates suggest that state-of-the-art experiments in both superconducting circuits and ultracold atomic gases are promising platforms for experimentally investigating the relationship between measurement and chaos, and are capable of observing chaotic dynamics deep within the quantum regime. Experimentally, one possible approach to infer the degree of chaos would be time series analysis~\cite{Kantz:1997,Trostel:2018}.  This requires acquisition of large data sets, which is possible in experiments, but  computationally expensive for large-scale quantum simulations.  Theoretically, it is much simpler to calculate Lyapunov exponents directly.

Although our initial investigations have revealed that this adaptive measurement scheme shows promise, our model did not include the effect of detection inefficiency. Detection inefficiency could affect both the emergence of chaotic dynamics and the effectiveness of our adaptive measurement protocol. For the quantum Duffing oscillator, numerical simulations have shown positive Lyapunov exponents with measurement efficiencies as low as 20\%~\cite{Ralph:2017a}. These Lyapunov exponents were also shown to be robust to small errors in the system parameters.  Measurement efficiencies as high as 80\% have been reported in recent superconducting circuit experiments~\cite{Eddins:2018}.  Similar detection efficiencies are possible in BEC systems at the cost of introducing heating, the effects of which would require further investigation.

In addition to perfect detection efficiency, our model assumes that the underlying estimate of the system state used to effect feedback (through the choice of quadrature measurement angle) precisely corresponds to the underlying system state. Although conditional master equations are known to be robust to imperfections in such estimates, which arise due to imperfect estimates of the model parameters, technical noise sources, and time delays, relaxing this assumption through a system-filter separation would provide crucial detail needed for the experimental realization of our adaptive measurement protocol~\cite{Szigeti:2013}. 
This work has focused on the control of chaos with continuous measurement in a \emph{single-particle} system. \emph{Many-body} quantum chaos is a growing research field, due to its potential connections to random unitaries~\cite{Elben:2018}, information scrambling and holographic duality ~\cite{Maldacena:2016,Swingle:2016,Meier:2017}, nonequilibrium thermodynamics~\cite{Neill:2016aa}, and even quantum sensing~\cite{Fiderer:2018}. Whether measurement can be used to meaningfully control chaos in many-body quantum systems is an intriguing question that warrants further investigation. 

\section{Conclusion}

In this work we have shown that the degree of chaos in a quantum Duffing oscillator can be controlled by applying real-time state-dependent feedback via an adaptive measurement technique.
The underlying mechanism for this control is the rate at which the measurement backaction destroys interference fringes in the state's Wigner function. By adaptively choosing measurements that are more (less) destructive, the dynamics more (less) closely resemble the corresponding classical trajectory, thereby enhancing (suppressing) chaos. Using this adaptive measurement technique, we have shown that the presence of chaos can be pushed further into the quantum regime. This regime is more easily accessible for certain experimental setups, potentially enabling new, detailed studies into the emergence of chaos in quantum systems. 

\section*{Acknowledgements}
The authors would like to thank A.~Pattanayak and S.~Greenfield for thoughtful discussions. The authors would also like to thank P.J.~Everitt for discussions on the experimental details and parameters.

J.K.E. acknowledges the support of an Australian Government Research Training Program (RTP) Scholarship and the hospitality of the Centre for Quantum Dynamics at Griffith University, where part of this work was completed. J.K.E. and A.R.R.C acknowledge support by the Australian Research Council (ARC) Centre of Excellence for Quantum Computation and Communication Technology (project CE110001027). S.S.S. received funding from ARC projects DP160104965 and DP150100356.
This research was undertaken with the assistance of resources and services from the National Computational Infrastructure (NCI), which is supported by the Australian Government.

\appendix

\section{Numerical simulation} \label{sec_appendix}

We numerically simulated the SSE~(\ref{eq:SSEITO}) on a finite subspace of $N$ 
energy eigenstates of the harmonic oscillator by using the software package XMDS2~\cite{Dennis:2013}. That is, we write the conditional state as $| \psi \rangle = \sum_{n=0}^{N-1} C_n(t) \vert n \rangle$ and numerically solve for the dynamics of the coefficients $C_n(t)$, governed by the set of Stratonovich stochastic differential equations 
\begin{widetext}
\begin{align}
	\label{eq:strat}
	\mathrm{d}C_n	&= -i \Big[ \tfrac{\beta^2}{4}\sqrt{(n+1)(n+2)(n+3)(n+4)} C_{n+4} 
				+  \sqrt{(n+1)(n+2)} ( \tfrac{\beta^2}{4} (4n+6) -\tfrac{1}{2}(1+i\Gamma ) ) C_{n+2}\notag \\ 
				&- \tfrac{g}{\sqrt{2} \beta} \cos{(\Omega t)}\sqrt{n+1} C_{n+1} 
				+ \tfrac{\beta^2}{4} (6n^2+6n+3)  C_n 
				- \tfrac{g}{\sqrt{2} \beta}\cos{(\Omega t)}\sqrt{n}  C_{n-1}\notag \\ 
				&+ \sqrt{n(n-1)} (\tfrac{\beta^2}{4} (4n-2) - \tfrac{1}{2} (1 - i\Gamma) ) C_{n-2}
				+ \tfrac{\beta^2}{4} \sqrt{n(n-1)(n-2)(n-3)} C_{n-4} \Big]  \mathrm{d}t \notag \\ 
				&- n\Gamma C_n \mathrm{d}t 
				- e^{2i\phi} \Gamma \sqrt{(n+1)(n+1)}C_{n+2} \mathrm{d}t
				 + 2\Gamma \left(\langle \hat{a}^\dagger \rangle + \langle \hat{a} \rangle ~e^{2i\phi} \right) \sqrt{n+1} C_{n+1} \mathrm{d}t \notag \\
				&+ \sqrt{2\Gamma} \sqrt{n+1} C_{n+1}~ e^{i\phi} \circ \mathrm{d}W,
\end{align}
\end{widetext}

where $\langle \hat{a} \rangle = \sum_{n=0}^{N-2} \sqrt{n+1} C_n^* C_{n+1}$ and $C_n = 0$ for all $n \geq N$. %(the 
For our simulations, we use $N = 64$ basis states, a large enough number such that $\vert C_{N-4} \vert^2+ \vert C_{N-3}\vert^2 + \vert C_{N-2} \vert^2 + \vert C_{N-1} \vert^2 < 10^{-4}$ at all times, while still small enough to be numerically tractable.

For the adaptive protocol, we calculate the probability distribution for a number of quadratures, this is given by
\begin{eqnarray}
\label{eq:prob2}
P_{X_\phi}&=& \vert\langle X_\phi | \psi\rangle\vert^2 \notag \\
&=& \left|\sum_n C_n \psi_n(x) e^{-i n \phi}\right|^2,
\end{eqnarray}
where $\psi_n(x)$ are the Hermite-Gauss functions:
\begin{eqnarray}
\psi_n(x)&=& (2^n n! \sqrt{\pi})^{-1/2} e^{-x^2/2}~ \mathrm{H}_n(x),
\end{eqnarray}
and $\mathrm{H}_n(x)$ are Hermite polynomials.

We use a grid-based search algorithm to determine the optimum measurement phase for each time step. To do this, we use a finite-difference method to calculate the derivative of the probability distribution~(\ref{eq:prob}) for an equidistant grid of LO angles $\phi \in [0,\pi]$, allowing the number of peaks in the distribution to be calculated. The angle $\theta_\text{max}$ corresponding to the maximum number of peaks gives an axis perpendicular to the interference fringes ($\theta_f + \pi/2$). To enhance chaos we adjust the LO phase to $\phi = \theta_f$ (parallel to fringes), whereas to suppress chaos we choose $\phi = \theta_f + \pi/2$ (perpendicular to fringes) for the next integration step of Eq.~(\ref{eq:strat}). 
In order for this grid-based search method to be effective, the grid of LO angles used needs to be of sufficiently high resolution. We found that, when suppressing chaos ($\phi = \theta_f + \pi/2$), a grid of 32 angles was required, whereas for enhancing chaos ($\phi = \theta_f$), a coarser grid of 8 angles was sufficient.

%We quantify the degree of chaos in our system by computing the quantum Lyapunov exponent as in Ref.~\cite{Eastman:2017}, which is based on an adaptation of the usual classical procedure~\cite{Wolf:1985}.
%The quantum Lyapunov exponent is defined as $\lambda =\lim_{t \to \infty} \lim_{d_0 \to 0} \log{\left(d_t/d_0\right)}/t$. Here $d_t=\sqrt{\Delta Q(t)^2 + \Delta P(t)^2}$ is the dimensionless phase-space distance between two quantum trajectories with differences in the average position and average momentum of the two trajectories given by $\Delta Q(t) = \langle \hat{Q}_{1} \rangle - \langle \hat{Q}_{2} \rangle$ and $\Delta P(t) = \langle \hat{P}_{1} \rangle - \langle \hat{P}_{2} \rangle$, respectively. The two quantum trajectories are initially prepared in coherent states displaced (in phase space) from each other by a small distance $d_0 = d_{t=0}$ (i.e. $|\alpha_1 \rangle = |\alpha\rangle$ and $|\alpha_2 \rangle = |\alpha + d_0\rangle$), and then evolved stochastically via Eq.~(\ref{eq:SSEITO}) under the same noise realization, which corresponds to the same measurement record. 
%

We quantify the degree of chaos in our system by computing the quantum Lyapunov exponent as in Ref.~\cite{Eastman:2017}, which is based on an adaptation of the usual classical procedure~\cite{Wolf:1985}. For our numerical calculations, one of the trajectories is periodically reset towards the other one to remain within the linear regime and $\log{\left(d_t/d_0\right)}$, calculated before every reset, is averaged over time. The perturbed trajectory after the reset is a displaced version of the trajectory of interest. The displacement is given by the initial distance $d_0$ in phase space, in the direction of expansion. The perturbed trajectory becomes $\ket{\psi_2} = D(\alpha) \ket{\psi_1}$, where $D(\alpha)$ is the displacement operator and $\alpha = d_0[(\langle \hat{Q}_{2} \rangle + i\langle\hat{P}_{2} \rangle) - ( \langle \hat{Q}_{1} \rangle + i\langle\hat{P}_{1} \rangle)] / (d_t \beta)$ is the displacement in the direction of expansion.

The simulations are run over $10,000$ cycles of the driving term ($t = 10^4/\Omega$) for both the adaptive- and the fixed -LO cases, and the final Lyapunov exponent is averaged over multiple realizations (10 runs) of the stochastic noise.

\end{document}